\documentclass[fleqn,usenatbib]{mnras}

\usepackage{newtxtext,newtxmath}

\usepackage[T1]{fontenc}
\usepackage{ae,aecompl}
\usepackage{threeparttable}
\usepackage{graphicx}	
\usepackage{amsmath}	
\usepackage{amssymb}	

\title[Exoplanet Candidates in the Galactic Bulge]{Search for Exoplanetary Transits in the Galactic Bulge}

\author[C.C. Cort\'es et al.]{
C.C.Cort\'es,$^{1}$\thanks{E-mail: cacortes@udec.cl} 
D. Minniti$^{2,3,4}$ , and
S. Villanova$^{1}$
\\
$^{1}$Departamento de Astronom\'ia, Casilla 160-C, Universidad de
  Concepci\'on, Concepci\'on, Chile.\\
$^{2}$Departamento de F\'isica, Facultad de Ciencias Exactas, Universidad Andr\'es Bello, Av. Fernandez Concha 700, Las Condes, Santiago, Chile.\\
$^{3}$Instituto Milenio De Astrof\'isica, Santiago, Chile.\\
$^{4}$Vatican Observatory, V00120 Vatican City State, Italy.
}
\date{Accepted 2018 November 21. Received 2018 November 20; in original form 2018 August 29}

\pubyear{2018}

\begin{document}
\label{firstpage}
\pagerange{\pageref{firstpage}--\pageref{lastpage}}
\maketitle

\begin{abstract}
A search for extrasolar planetary transits using the extended Kepler mission (K2) campaigns 9 and 11 revealed five new candidates towards the Galactic bulge.
The stars EPIC 224439122, 224560837, 227560005, 230778501 and 231635524 are found to have low amplitude transits consistent with extrasolar planets, with periods $P= 35.1695, 3.6390, 12.4224, 17.9856$, and $5.8824$ days, respectively.
The K2 data and existing optical photometry are combined with the multi-band near-IR photometry of the VVV survey and 2MASS in order to measure accurate physical parameters for the host stars.
We then measure the radii of the new planet candidates from the K2 transit light curves and also estimate their masses using mass-radius relations, concluding that two of these candidates could be low mass planets, and three could be giant gaseous planets.
\end{abstract}



\section{Introduction}
The Kepler mission \citep{Borucky2010,Borucki2011} was a clear success and a revolution for extrasolar planet studies. The main mission lasted four years and the data collected is still producing extrasolar planets, that are now counted by the thousands.
The extended Kepler mission called K2 consisted of several campaigns, with multiple fields observed along the ecliptic plane since 2014. 
Among a variety of studies, K2 has discovered more than a hundred transiting extrasolar planets up to now \citep{Montet2015, Schlieder2016, Eylen2016, Johnson2016, Adams2016, Sinukoff2016, Barros2016, Pope2016, Dressing2017a, Dressing2017b, Petigura2018, Wittenmyer2018, Mayo2018, Yu2018, Crossfield2018, Livingston2018}.

We are interested here in the data from campaigns 9 and 11 (hereafter K2C9 and K2C11), that observed the Milky Way bulge.   
This is a crowded and reddened region of our Galaxy, but of great interest, because it overlaps with our ongoing VVV survey, that has been mapping the whole bulge in the near-IR since 2010 \citep{Minniti2010,Saito2012}. The challenge is the large Kepler 4 arcsec pixel scale, as discussed extensively elsewhere by \citet{Henderson2016} and \citet{Zhu2017}. We use our higher resolution VVV images (with 0.3''/pixel scale) in order to weed out bad candidates (usually blended objects). In particular, \citet{Henderson2016} describe in detail the goals, difficulties, and procedures of the K2C9.

We started a project to detect and study new transiting exoplanets in the Galactic Bulge, using K2 mission and VVV survey data.
We report here the discovery and characterization of five exoplanetary candidates in the bulge.
This paper is organized as follows, in Section ~\ref{sec:data} we present the K2 photometry for the campaign 9 and 11 and the VVV survey.
Section ~\ref{sec:search} discusses our search. In Section ~\ref{sec:stellar} and ~\ref{sec:planet}
we give the physical parameters for the sample stars and planets, respectively.
The discoveries are discussed in turn in Section ~\ref{sec:discussion}.
Finally, Section ~\ref{sec:conclusion} outlines our main conclusions.


\begin{figure*}
	\includegraphics[width=14cm,height=7cm]{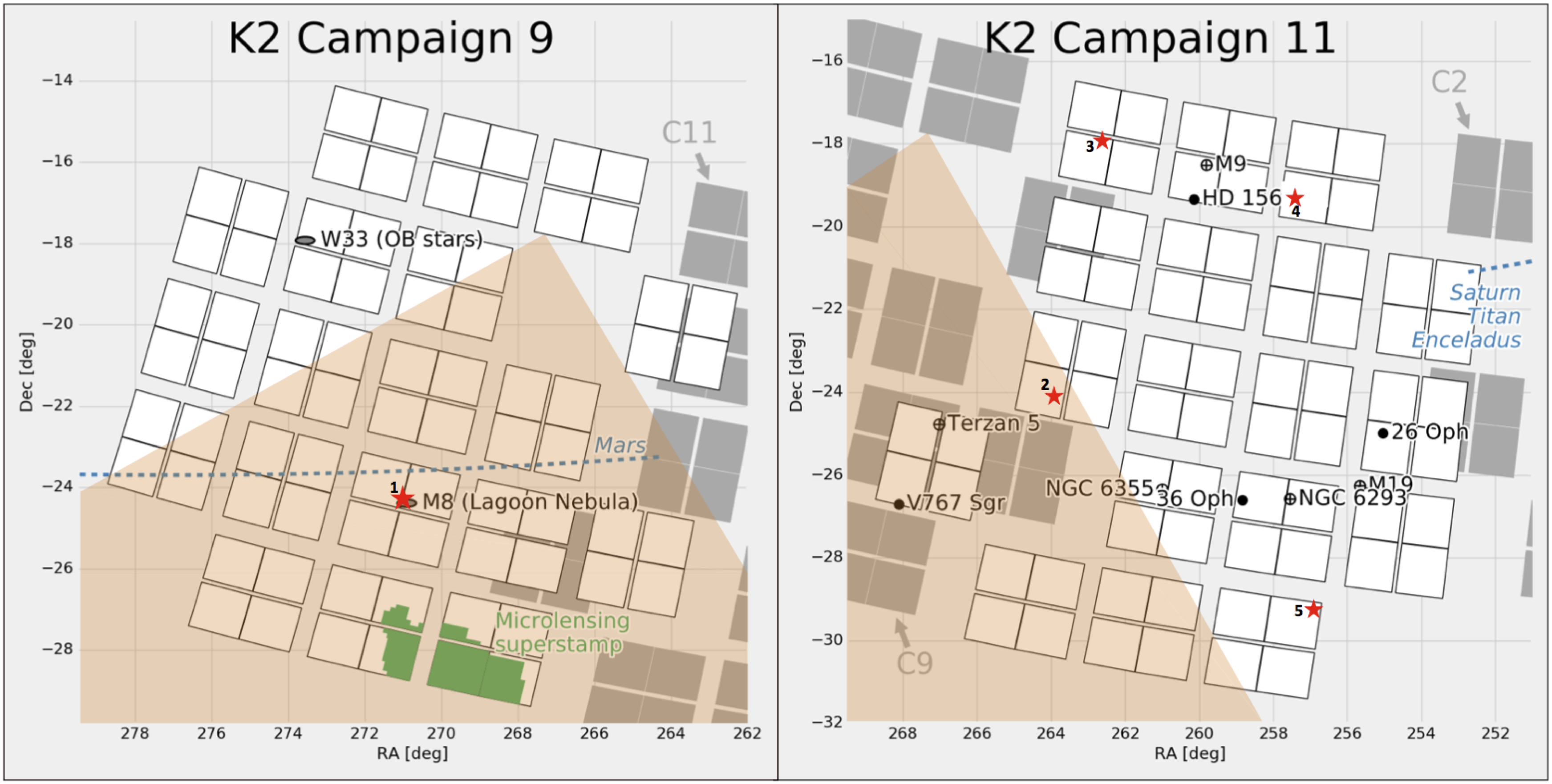}
    \caption{The figure shows the two K2 campaign fields: campaign 9 (left) and campaign 11 (right), in which we perform the search for extrasolar planets. The red stars represent the position of the exoplanet candidates that we found in our study with their respective numbers (see table~\ref{tab:table_planet}). The green area in campaign 9 represents the microlensing super apertures. The overlapped coloured region corresponds to the region mapped by the VVV survey. This plot is taken from the Kepler $\&$ K2 website.}
    \label{fig:campaign}
    \centering
\end{figure*}


\section{Data}
\label{sec:data}

\subsection{K2 Photometry}
\label{sec:k2}
For our study, we used the K2 database, which provides high-precision photometry on the 1 and 30 minute timescales. The Kepler magnitude ($K_{p}$) refers to an AB magnitude, ranging from 425 to 900 nm.
The Kepler photometer consists in multi CCD modules, and each module covers 5 square degrees on the sky. The observations of K2 consist of a series of observation field ``Campaigns" distributed in the plane of the ecliptic.

The campaign 9 consisting of 19 CCD modules (Figure~\ref{fig:campaign}, left) covered part of the Galactic Bulge and was dedicated to a microlensing study. In order to increase the data storage, the campaign 9 was split into two parts (campaign 9a and 9b), with a three-day gap from May 19 to May 22, 2016. 
The campaign 9a is centred at RA=270.3544823 degrees, DEC=-21.7798098 degrees and was observed between April 22 and May 19, 2016. The campaign 9b that was observed between May 22 and July 1, 2016, and is centred at RA=270.3543824 degrees, DEC=-21.7804700 degrees. For our study, we considered only targets from the microlensing super apertures (green area in the Figure~\ref{fig:campaign}, left ), in which  3.3 million pixels were dedicated on five CCD channels.

The campaign 11 consists of 18 CCD modules, due to the loss of CCD module 4. This campaign is centred at RA=260.3880064 degrees, DEC=-23.9759578 degrees (Figure~\ref{fig:campaign}, right) and covers part of the Galactic Bulge. This campaign was split into two parts with a three-day gap from October 18 to October 21, 2016: the campaign 11a that was observed between September and October 2016 during 23 days and the campaign 11b was observed between October and December 2016 during 48 days.

\subsection{VVV survey}
The VISTA Variables in the V\'ia L\'actea (VVV) survey \citep{Minniti2010,Saito2012}
 that covers a bulge area of  300 square degrees between $-10^{\circ}$ < l < $+10^{\circ}$ and $-10^{\circ}$ < b < $+5^{\circ}$ divided in 196 tiles.

This survey provides near-infrared photometry in five broad-band filters: Z (0.87 $\mu m$), Y (1.02 $\mu m$), J (1.25 $\mu m$), H ($\mu m$) and, $K_{s}$ (2.14 $\mu m$).
We used the VVV photometric catalogue that was obtained from the Cambridge Astronomical Survey Unit (CASU) \footnote{http://casu.ast.cam.ac.uk/vistasp/} in different tiles in the galactic bulge. The K2-VVV areas of overlap are shown in Figure~\ref{fig:campaign}.


\section{Search for Exoplanetary transits}
\label{sec:search}
For the exoplanet candidates search, we used 875 light curves from the campaign 9 and 13.607 light curves from the campaign 11, that were extracted by \citet{Vanderburg2014}.
Both campaigns were split into two parts, therefore we normalized the flux of the light curve for part a and b (see section~\ref{sec:k2}) using a cubic spline function. We choose the order of the polynomials according to the light curve. After the fitting, we removed upwards outliers, which are caused by cosmic rays or asteroids and also, we removed the downward outliers making sure that the transit was not removed. 
After flattening the light curves, we calculated a Box Least Squares (BLS)\footnote{https://github.com/dfm/python-bls} periodogram \citet{kovacs02}, to detect a periodic signal. We used the definition of \citet{Vanderburg2016} to perform the period search ranging from 2.4 hours to half the length of the campaign and the spacing between periods expressed as:
\begin{equation*}
\Delta P = P \frac{D}{N  \times T_{tot}}
\end{equation*}
where $\Delta P$ is the spacing between periods, P is the period tested, D is the transit duration at that period, N is an oversampling factor and $T_{tot}$ is the total duration of the campaign.

After this process, we cleaned our catalogue by applying some restrictions. From the analysis described by \citet{Vanderburg2016}, we considered targets that in the BLS periodogram have at least one peak with $S/N > 9$. Also, we eliminate objects whose duration is greater than 20\% of the detected period, and we considered only detections that have two or more transit events.

Even when an object passes these tests, there is the possibility that it is a false positive. For this reason, we performed a visual inspection to discard obvious false positives such as spurious detections, eclipsing binaries and any other astrophysical false positives.

Additionally, we take advantage of the near-infrared data from the VVV survey that overlapped with the K2 data (see Figure~\ref{fig:campaign}), to discard false positives, specially blended objects, because with this photometry we can constrain the contaminant stars with a different colour than the target \citep{Fressin2012}.

Our final catalogue contains five planet candidates (see table~\ref{tab:table_planet}). The planetary parameters estimation is explained in section ~\ref{sec:planet}.


\section{Stellar Properties}
\label{sec:stellar}
The stellar parameters of the host stars of our exoplanet candidates are summarized in Table~\ref{tab:table_stellar}. Based on previous studies, the host star 224439122 was catalogued as a variable Weak T tau \citep{Prisinzano2012} with a period 5.8775 days \citep{Henderson2012}.

The stellar parameters for our exoplanet candidates were estimated through Gaia DR2 \citep{Andrae2018}, whose information was extracted from Gaia Archive\footnote{https://gea.esac.esa.int/archive/} \citep{Gaia2016,Gaia2018}. 
The radius of the host star 231635524 cannot be derived from data in Gaia DR2, because stars with fractional parallax uncertainties greater than 20 percent are not reliably inverted to yield distances. This particularly affects distant and/or faint stars \citep{Andrae2018}. For the host star 231635524, the fractional parallax uncertainty is 450 percent (\citealt{Jones2018} infer a distance in excess of 10 kpc with large errors). It is most likely this host star is at least a giant, since this star is too bright to be a dwarf, given the likely very long distance. 

We classify our host stars by calculating the reduced proper motion, after properly correcting for extinction. For the passbands in the VVV survey, we measured the extinction using the reddening maps of \citet{Gonzalez2012} by the tool BEAM (Bulge Extinction And Metallicity) calculator\footnote{http://mill.astro.puc.cl/BEAM/calculator.php} using the \citet{Cardelli1989} extinction law. 
In the case of filters in the 2MASS catalogue, we used \citet{Schlegel1998} maps through the tool Galactic DUST Reddening \& Extinction\footnote{https://irsa.ipac.caltech.edu/applications/DUST/}. The photometric parameters are summarized in Table~\ref{tab:photo}.

With the proper motion of Gaia DR2 catalogue, we calculated the reduced proper motion of our host stars, through the equation defined as:
\begin{equation}
H_{J}= J + 5 \log \mu
\end{equation}
where J is the J-band magnitude and $\mu$ is the total proper motion. With the criteria defined by \citet{Rojas2014}:
\begin{equation}
H^{dwarf}_J > H^{*}_J = 68.5 (J-K_{s}) - 50.7
\end{equation}
we classify the host star like dwarf or giant, where $H^{*}_J$ is the dwarf/giant discriminator.

\section{Planetary Parameters}

\label{sec:planet}

We have found five planet candidates, with the period and depth calculated from the output of the BLS algorithm. Assuming that the orbit is circular, we modeled the transit time, the period, the planetary to stellar radius ratio ($R_{p}/R_{\star}$), the semi-major axis normalized to the stellar radius ($a/R_{\star}$) and the inclination, through a transit model using the BAsic Transit Model cAlculatioN (BATMAN) \footnote{http://astro.uchicago.edu/~kreidberg/batman/tutorial.html} Python package \citep{BATMAN}. For the development of our model, we used the quadratic limb darkening law \citep{Kopal1950} with the coefficient estimated by \citet{BATMAN}.

In order to take into account the smearing effect of the 30 min cadence of the K2 data \citep{Kipping2010}, we used the supersampling provided by BATMAN, that consists in calculating the average value of the light curve from the evenly spaced samples during an exposure.

After that, we measured the transit parameters and their uncertainties of this model using emcee\footnote{http://dfm.io/emcee/current/} Python package \citep{emcee}, which is an implementation of the affine-invariant ensemble sampler for Markov Chain Monte Carlo (MCMC) \citep{Goodman2010}. We implement the same uncertainties to each flux, because the flux error was not calculated in the K2 data reduction process.

We estimated the mass of the planet candidates through Forecaster\footnote{https://github.com/chenjj2/forecaster} Python package developed by \citet{Chen2017}, which uses a probabilistic mass-radius relationship. With this code it is possible to predict the mass of the candidates based on a given radius measured previously.

\begin{figure*}
	\includegraphics [width=11cm,height=9cm]{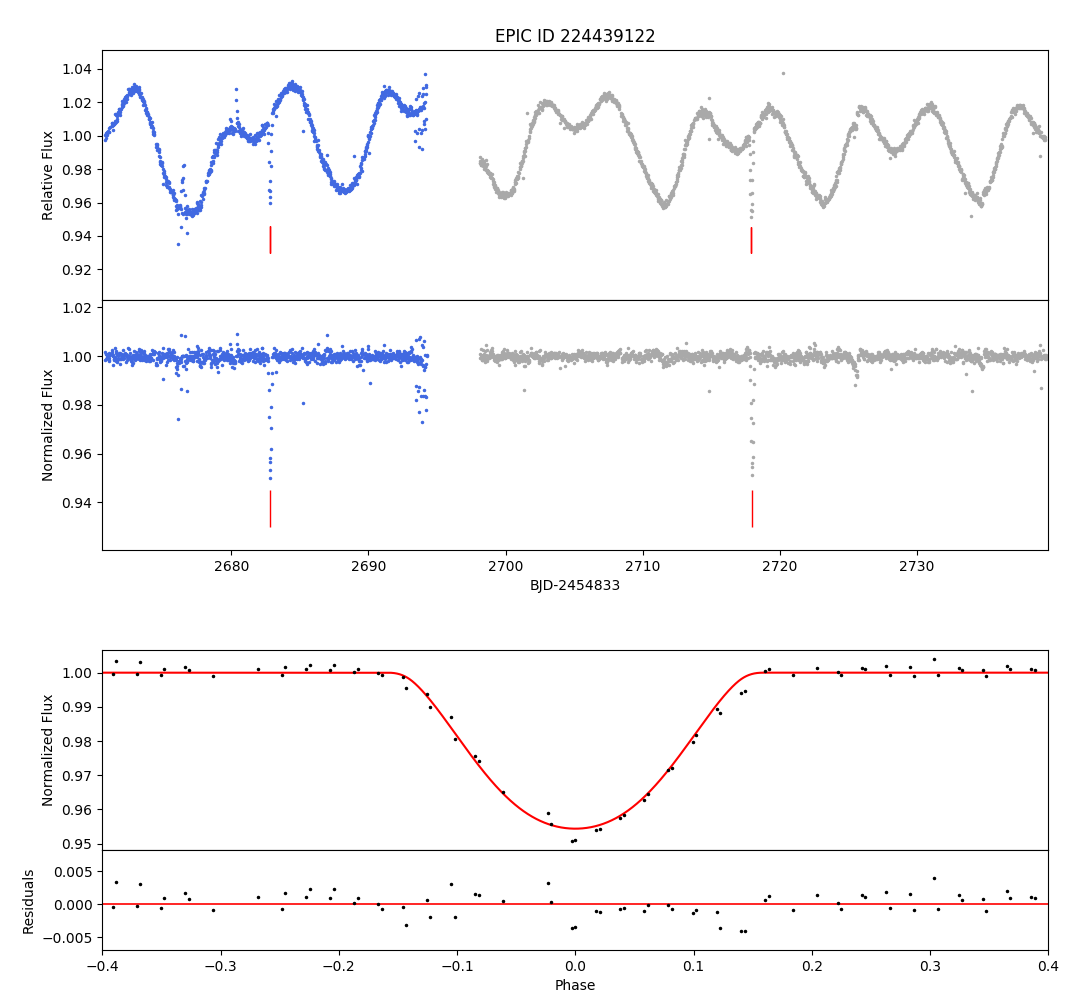}
    \caption{Top: Light curve from the campaign 9 of the exoplanet candidate 224439122 orbiting a variable star. As we mention in the section~\ref{sec:k2} the campaign was split into two parts: the blue points indicate c9a and the gray points indicate c9b. The red lines show two transit times. Middle: Flattened light curve of EPIC ID 224439122 (see section~\ref{sec:search}). Bottom: Phase folding of the normalized light curve and residuals. The black points mark the K2 data, and the red line marks the best-fitting transit model. This candidate could be a gaseous giant with a period of 35.1695 days.}
    \label{fig:light_curve_1}
    \centering
\end{figure*}

\begin{figure}
	\includegraphics [width=9cm,height=8cm]{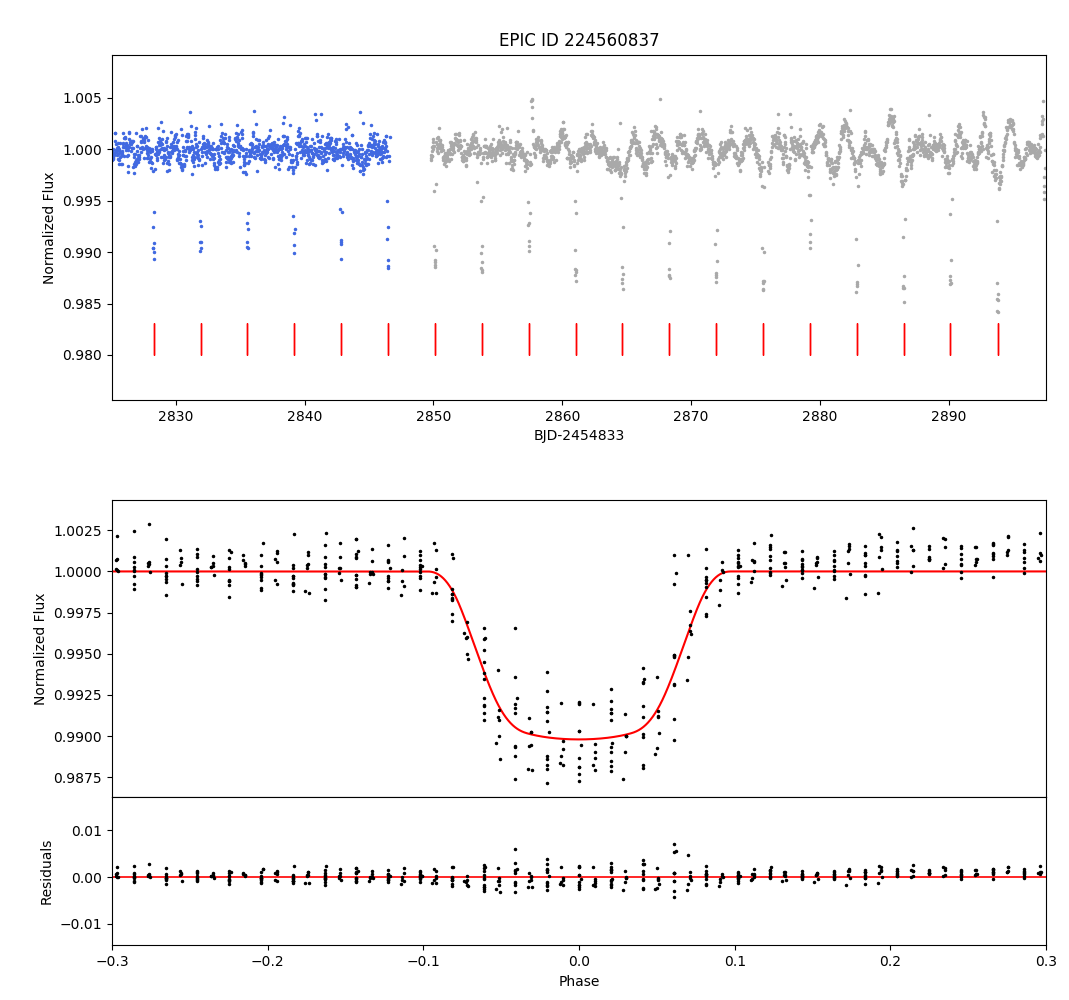}
    \caption{Top: Flattened light curve from the campaign 11 of the exoplanet candidate 224560837. The blue points indicate c11a and the gray points indicate c11b. The red lines show nineteen transit times. Bottom: Phase folding of the normalized light curve and residuals. The black points mark the K2 data, and the red line marks the best-fitting transit model. This is a possible gaseous giant in a 3.6390 days orbit.}
    \label{fig:light_curve_2}
    \centering
\end{figure}

\begin{figure}
	\includegraphics [width=9cm,height=8cm]{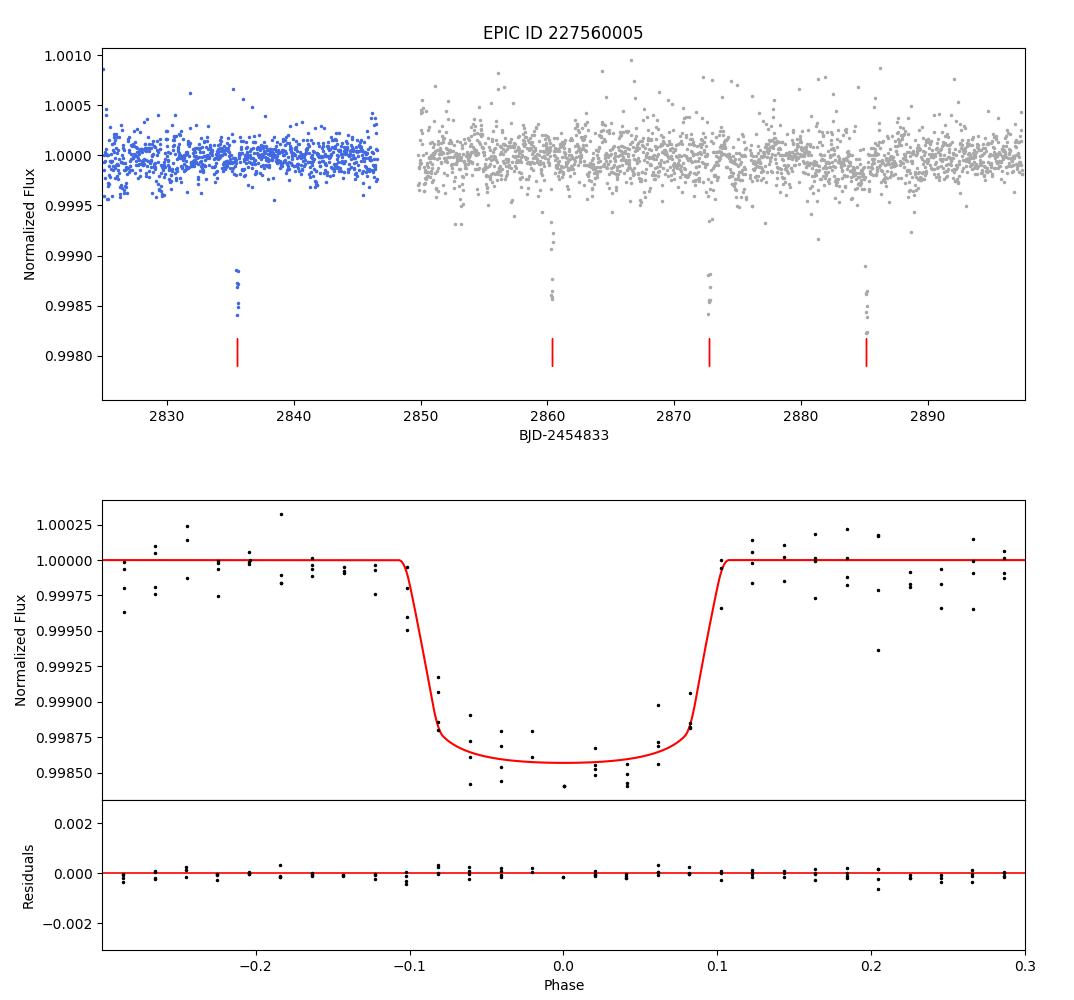}
    \caption{ Light curve from the campaign 11 of the exoplanet candidate 227560005. See Figure~\ref{fig:light_curve_2} caption. This target could be a low mass planet with period P=12.4224 days and four transit events.}
    \label{fig:light_curve_3}
    \centering
\end{figure}

\begin{figure}
	\includegraphics [width=9cm,height=8cm]{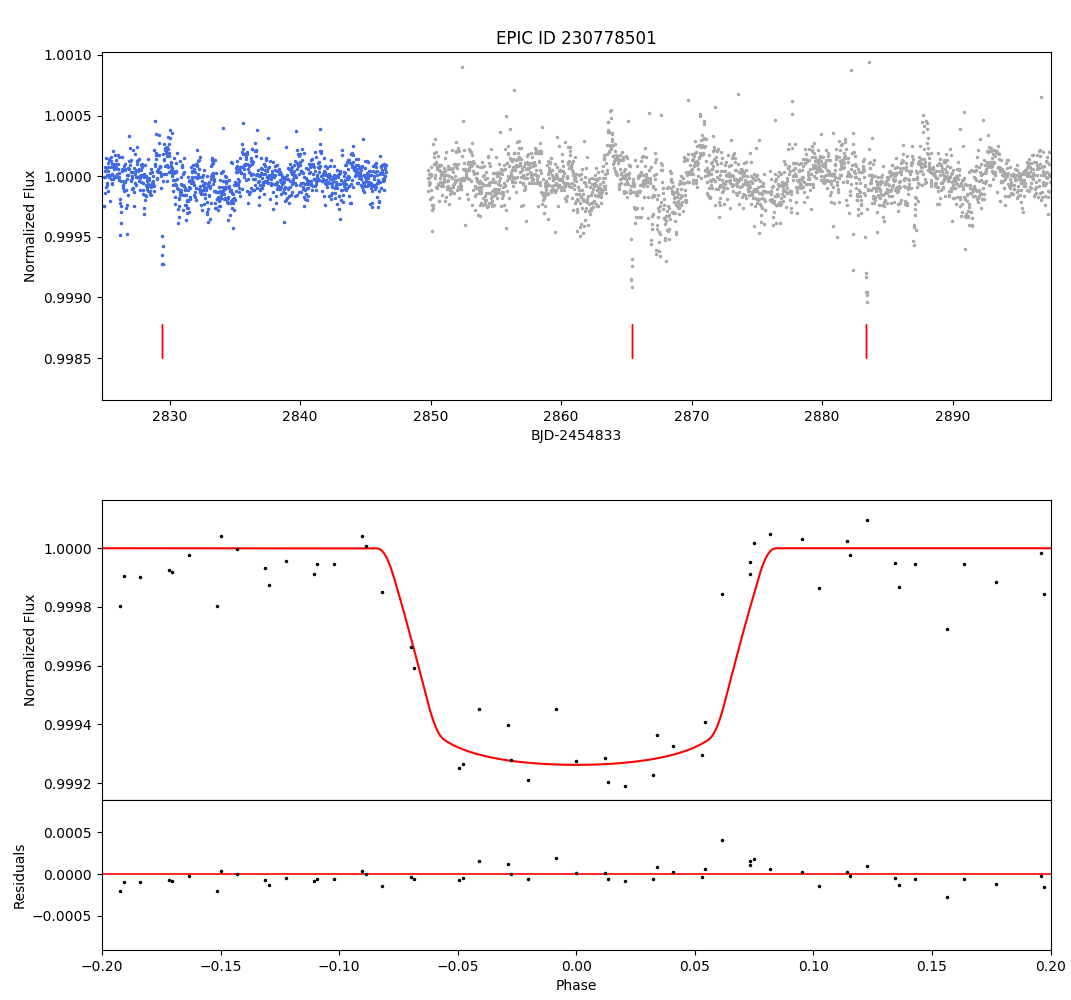}
    \caption{Light curve from the campaign 11 of the exoplanet candidate 230778501. See Figure ~\ref{fig:light_curve_2} caption. This target could be a low mass planet with a period of 17.9856 days and three transit events.}
    \label{fig:light_curve_4}
    \centering
\end{figure}

\begin{figure}
	\includegraphics [width=9cm,height=8cm]{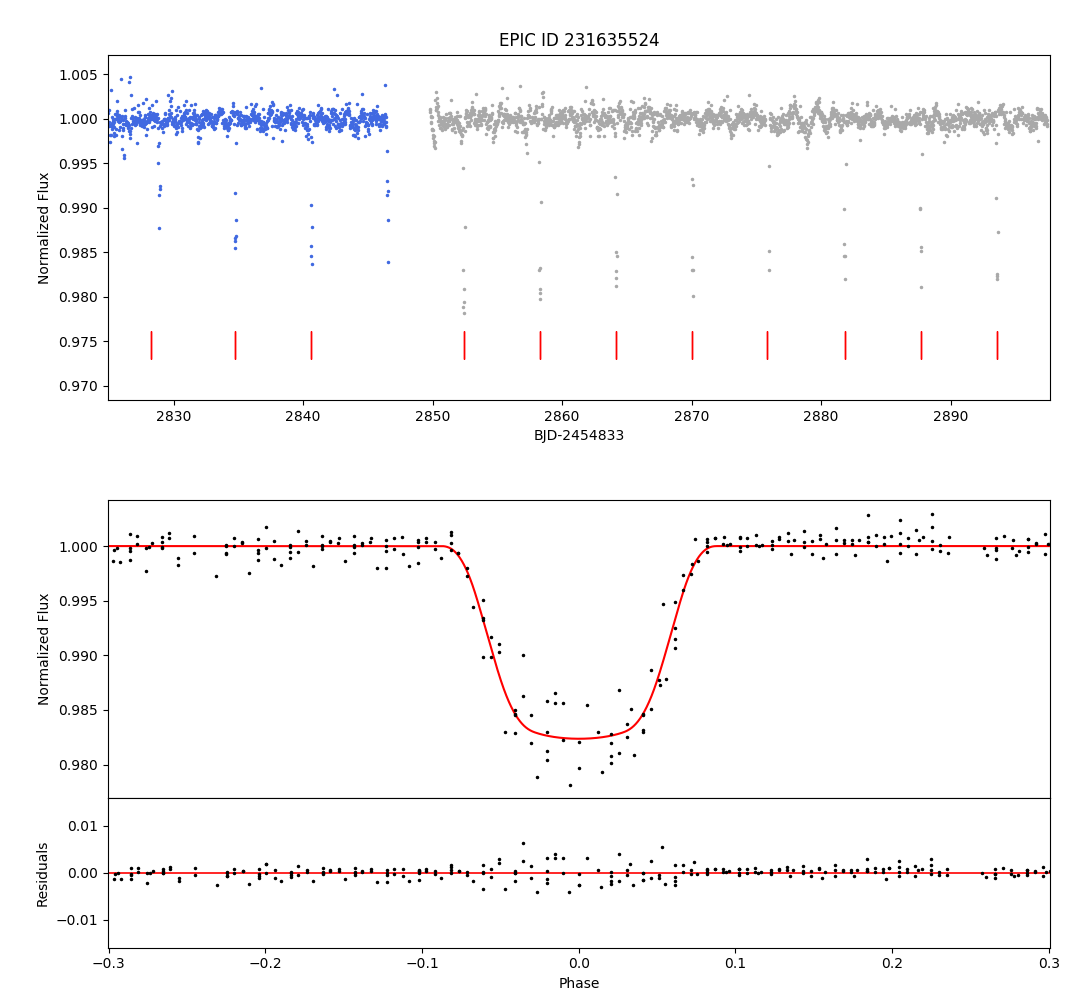}
    \caption{Light curve from the campaign 11 of the exoplanet candidate 231635524. See Figure ~\ref{fig:light_curve_2} caption. This target could be a gaseous giant with a period P=5.8824 days and eleven transit events.}
    \label{fig:light_curve_5}
    \centering
\end{figure}

The estimation of the planetary parameters is summarized in Table~\ref{tab:table_planet}. The exoplanet candidates with the fitting model are shown in Figure~\ref{fig:light_curve_1} to Figure~\ref{fig:light_curve_5}.

\section{Discussion}
\label{sec:discussion}
In this work, we present the discovery of five exoplanet candidates, which were detected with the transit method using K2 photometry. One of the parameters that we can obtain with this method is the planetary radius. To determine this parameter, we need the stellar radius, which was calculated photometrically (see Section~\ref{sec:stellar}). To definitely classify these candidates is it is necessary to estimate their masses, which are not possible to measure using this technique. Clearly, spectroscopic observations are needed for these candidates.
Therefore, we predict the mass through the code Forecaster (see Section~\ref{sec:planet}), which uses a mass-radius relationship. Even though these parameters are not highly accurate, they represent a good initial estimate to perform our analysis and give a preliminary idea about the nature of our candidates.

We compared our candidates with the mass-density relationship proposed by \citet{Hatzes2015}. The relationship is based on the inflections in the mass-density diagram, and shows three regions. The regions are low mass planets, giant gaseous planets and stellar objects.
Then, we proceed to analyze each of our candidates:
\\
\\
\textbf{EPIC 224439122b} is a candidate exoplanet orbiting the host star located in NGC~6530 (open cluster), classified as a variable Weak T Tau star with spectral type M0-M1 V \citep{Prisinzano2012} with a period 5.8775 days \citep{Henderson2012}.
This extrasolar planet candidate has two transits and has a period 35.1695 days, indicating that it could be a warm Jupiter (the orbital period between 10 and 100 days). Also, this is the largest candidate in our sample with R=48.1$R_{\earth}$ and an estimated mass of M=438.0$M_{J}$, implying that it could be a stellar object ($M>60M_{J}$, \citealt{Hatzes2015}). 
As the largest planets known have radii of $\sim 20 R_{\earth}$, this radius seems to be too large for an exoplanet.
Although this candidate has a depth of $\sim 5\%$ and a large mass, we consider that this object passed the test described in Section \ref{sec:search} to discard false positive. Also, as we mentioned above, we do not measure the mass and the stellar parameters. Therefore these are not entirely reliable.
\\
\textbf{EPIC 224560837b} is a candidate with nineteen transit events with a period of 3.6390 days, which is within the definition of hot Jupiters (the orbital period between 1 and 10 days). This candidate has a radius of 30.6$R_{\earth}$ and an estimated mass of 260.8$M_{J}$ indicating that it could be a stellar object ($M>60M_{J}$, \citealt{Hatzes2015}). Despite this candidate is the second largest in our sample and according to the mass classification this could be a stellar object, we take into account that this target passed the test mentioned in Section \ref{sec:search} and the estimation of the stellar parameters and the planetary mass were not measured. For this reason, we do not discard the possibility that this object could be an extrasolar planet.
\\
\textbf{EPIC 227560005b} has a period of 12.4224 days. This candidate has four transits and is our smallest exoplanet candidate with a R=2.0$R_{\earth}$ and an estimated mass of 0.02$M_{J}$ indicating that it could be a low mass planet ($M<0.3M_{J}$, \citealt{Hatzes2015}). The same definition would apply for \textbf{EPIC 230778501b} that has three transits with a period of 17.9856 days. This candidate is the second smallest in our sample with R=2.2$R_{\earth}$ and an estimated mass of 0.02$M_{J}$.
\\
\textbf{EPIC 231635524b} is a candidate that has eleven transits with a period of 5.8824 days. According to the period, this could be a hot Jupiter. As we explained in Section \ref{sec:stellar}, the radius of the host star 231635524 is not available in Gaia DR2, and we infer that this star could be a giant. Therefore if 231635524 could be a giant and the planetary to stellar radius ratio is 0.132 (see Table \ref{tab:table_planet}) probably our candidate could be a giant gaseous planet.


\section{Conclusions}
\label{sec:conclusion}

We reported the discovery of five exoplanet candidates detected in the Galactic Bulge with K2 data with orbital periods between 3.6390 and 35.1695 days and planetary radii in the range of 2.0 to 48.1 $R_{\earth}$. These planet candidates orbit host stars with a range of magnitudes and temperatures (12.039 < $K_{p}$ < 16.072, and 4184 K < $T_{eff}$ < 4647 K).

Additionally, two of our candidates were classified as stellar objects (224439122 and 224560837) and two as low mass planets (227560005 and 230778501) according to \citet{Hatzes2015}, but we considered that 224439122 and 224560837 passed the test to discard false positive mentioned in Section~\ref{sec:search}, therefore there are the possibility that these targets could be planets. Due to the radius for the host star of the candidate 231635524 is not available in Gaia DR2, we can not estimate the planetary radius, and consequently, we can not predict the mass of the planet. Therefore, as we explained in Section \ref{sec:discussion}, we infer that the candidate 231635524 could be a giant gaseous planet. 

In addition, we emphasize that the stellar parameters were determined using photometry, and that the derived masses in particular are very uncertain. Therefore, we would like to encourage follow-up spectroscopic observations in order to confirm our exoplanet candidates and to refine their physical parameters.

\begin{table*}
   \centering
   \caption{Stellar parameters}
   \begin{threeparttable}[b]
	\label{tab:table_stellar}
	\begin{tabular}{cccccccccc} 
		\hline
        ID & EPIC ID & RA & DEC & $T_{eff}$ & $R_{\star}$ &  RPM $H_{J}$ & classification \\
        & & (J2000, h:m:s) & (J2000, d:m:s)  & (K) & ($R_{\sun}$) & & & \\
		\hline
        \hline
        & \hspace{-1.5cm}Campaign 9 \\
        \hline
		1 & 224439122 & 18:04:44.11 & -24:14:39.06 & $4348^{+106}_{-190}$\tnote{(1)} & $1.96^{+0.18}_{-0.09}$\tnote{(1)} & -2.60\tnote{(1)} & Dwarf\\
         \hline
        & \hspace{-1.3cm}Campaign 11 \\
        \hline
        2 & 224560837 & 17:35:13.92 & -24:01:42.35 & $4191^{+217}_{-199}$\tnote{(1)} & $2.75^{+0.28}_{-0.26}$\tnote{(1)} &  -1.09\tnote{(1)} & Dwarf \vspace{0.1cm} \\
        3 & 227560005 & 17:31:02.07	& -17:50:35.39 & $4184^{+124}_{-179}$\tnote{(1)} & $0.51^{+0.05}_{-0.03}$\tnote{(1)} &  4.20\tnote{(1)} & Dwarf \vspace{0.1cm} \\
        4 & 230778501 & 17:09:39.53	& -19:13:01.41 & $4647^{+141}_{-134}$\tnote{(1)} & $0.73^{+0.04}_{-0.04}$\tnote{(1)} &  3.24\tnote{(1)} & Dwarf \vspace{0.1cm} \\
        5 & 231635524 & 17:07:33.40	& -29:16:21.27 & $4322^{+277}_{-75}$\tnote{(1)} & - &  1.80\tnote{(1)} & Giant \vspace{0.1cm} \\
        \hline
	\end{tabular} 
     \textbf{Note.} RPM $H_{J}$ is the reduced proper motion of the star. RPM $H_{J}$ was calculated using the equation (1), with the proper motion from Gaia DR2 catalogue. In column 10, we classify the host star as giant and dwarf according to the criteria defined by \citet{Rojas2014} equation (2).  \\
     \textbf{References.}  (1) \citet{Gaia2016,Gaia2018}.
  \end{threeparttable}
\end{table*}

\begin{table*}
	\centering
	\caption{Photometric parameters}
    \begin{threeparttable}[b]
	\label{tab:photo}
	\begin{tabular}{cccccccccccccc} 
		\hline
		ID & EPIC ID & $K_{p}$ & U  & B & V & R & I & z & Y & J & H & $K_{s}$ \\
        & & (mag) & (mag) & (mag) & (mag) & (mag) & (mag) & (mag) & (mag) & (mag) & (mag) & (mag) \\
		\hline
        \hline
        & \hspace{-1.5cm}Campaign 9 \\
        \hline
		1 & 224439122 & 16.072& 17.445\tnote{(1)}& 17.445\tnote{(1)} & 16.145\tnote{(1)} & - & 14.560\tnote{(1)} & 14.132\tnote{(2)} & 13.786\tnote{(2)} & 10.395 \tnote{(2)} & 10.878\tnote{(2)} & 11.257 \tnote{(2)}\\
        \hline
        & \hspace{-1.3cm}Campaign 11 \\
        \hline
        2 & 224560837 & 14.931 & - & 16.443\tnote{(3)} & 15.174\tnote{(3)} & 14.723\tnote{(3)} & - & 13.068\tnote{(2)} & 12.828\tnote{(2)} & 11.525\tnote{(2)} & 11.514\tnote{(2)} & 11.481\tnote{(2)} \\
        3 & 227560005 & 12.039 & - & 14.132\tnote{(4)} & 12.706\tnote{(4)} & 12.088\tnote{(4)} & 11.402\tnote{(4)} & - & - &9.517\tnote{(5)} & 9.006\tnote{(5)} & 8.883\tnote{(5)}\\
        4 & 230778501 & 12.388 & - & 13.794\tnote{(3)} & 12.685\tnote{(3)} & 12.264\tnote{(3)} & 12.007\tnote{(3)} & - & - &10.351\tnote{(5)} & 9.879\tnote{(5)} & 9.760\tnote{(5)} \\
        5 & 231635524 & 14.749 & - & 14.940\tnote{(6)} & 14.030\tnote{(6)} & - & - & - & - & 12.311\tnote{(5)} & 11.787\tnote{(5)} & 11.709\tnote{(5)} \\
        \hline
	\end{tabular}
    \textbf{Note.} $K_{p}$ is the Kepler magnitude described in the section~\ref{sec:k2}. The filter J, H, and $K_{s}$ were corrected for extinction, as explained in section~\ref{sec:stellar}.  \\
    \textbf{References.} (1) \citet{Sung2000}; (2) VVV survey \citep{Minniti2017}; (3) \citet{Zacharias2012}; (4) APASS catalogue \citep{Henden2016}; (5) 2MASS All-Sky Survey \citep{Cutri2003}; (6) \citet{Zacharias2005}.
  \end{threeparttable}
\end{table*}

\begin{table*}
	\centering
	\caption{Planetary parameters}
    \begin{threeparttable}[b]
	\label{tab:table_planet}
	\begin{tabular}{cccccccccccc}
		\hline
		ID & EPIC ID & Period & Transit epoch $T_{0}$ & a/$R_{\star}$ & i & $\delta$ & $R_{P}/R_{\star}$ & $R_{P}$ & $M_{P}$ \\
           &    & (days) & BJD - 2454833 & & (deg) &($\%$) & & $(R_{\earth})$ & $(M_{\earth})$ \\
		\hline
        \hline
        & \hspace{-1.5cm}Campaign 9 \\
        \hline
        1 & 224439122  & 35.1695 & 2682.8129 & $34.09^{+0.16}_{-0.07}$  & $88.62^{+0.05}_{-0.02}$ & 5.05 & $0.225^{+0.004}_{-0.009}$ & $48.1^{+5.4}_{-0.5}$ & $139220.7^{+18450.3}_{-18814.8}$ \\ 
        \hline
        & \hspace{-1.3cm}Campaign 11 \\
        \hline
        2 & 224560837  &  3.6390 & 2828.2806 & $4.95^{+0.19}_{-0.21}$ & $80.48^{+0.54}_{-0.62}$ & 1.04 & $0.102^{+0.001}_{-0.001}$ & $30.6^{+3.5}_{-3.2}$ & $82890.3^{+14961.3}_{-14115.1}$  \vspace{0.1cm} \\
        3 & 227560005  & 12.4224   & 2835.5338 & $19.09^{+1.42}_{-1.23}$ & $88.68^{+0.30}_{-0.38}$ & 0.13 & $0.036^{+0.002}_{-0.003}$ & $2.0^{+0.3}_{-0.3}$ & $4.9^{+3.9}_{-2.1}$ \vspace{0.1cm} \\
        4 & 230778501  & 17.9856 & 2829.4447 & $27.78^{+1.37}_{-1.77}$ & $88.49^{+0.95}_{-0.67}$ & 0.07  & $0.027^{+0.011}_{-0.005}$ & $2.2^{+1.1}_{-0.5}$ & $6.1^{+4.7}_{-3.3}$ \vspace{0.1cm} \\
        5 & 231635524  &  5.8824 & 2828.8985 & $10.17^{+0.50}_{-0.35}$ & $85.71^{+0.35}_{-0.26}$ & 1.76 & $0.132^{+0.001}_{-0.001}$ & - & - \vspace{0.1cm} \\
        \hline
	\end{tabular}
    \textbf{Note.} The Sun and Earth units were obtained from the International Astronomical Union\citep{Prsa2016}. The planetary parameters were estimated in the section~\ref{sec:planet}. $T_{0}$ is the time of transit, $a/R_{\star}$ is the semi-major axis normalized to the stellar radius, $i$ is the inclination, $\delta$ is the transit depth, $R_{P}/R_{\star}$ the planetary to stellar radius ratio, $R_{P}$ is the planetary radius, calculated by multiplying the $R_{P}/R_{\star}$ values with the stellar radius and $M_{P}$ is the planet mass estimated using mass-radius relationship developed by \citet{Chen2017} through Forecaster Python package.
   \end{threeparttable}
\end{table*}


\section*{Acknowledgements}
We would like to thank the anonymous referee for careful review of this manuscript and for giving such valuable comments. C.C.C. is supported by CONICYT (Chile) throught Programa Nacional de Becas de Doctorado 2014 (CONICYT-PCHA/Doctorado Nacional/2014-21141084). S.V. and C.C.C. gratefully acknowledge the support provided by Fondecyt reg.n. 1170518. D.M. is supported by  FONDECYT Regular grant No. 1170121, by the BASAL Center for Astrophysics and Associated Technologies (CATA) through grant PFB-06, and the Ministry for the Economy, Development and Tourism, Programa Iniciativa Cientifica Milenio grant IC120009, awarded to the Millennium Institute of Astrophysics (MAS).
This paper includes data collected by the K2 mission. Funding for the K2 mission is provided by the NASA Science Mission directorate. We gratefully acknowledge use of data from the ESO Public Survey programme ID 179.B-2002 taken with the VISTA telescope, and data products from the Cambridge Astronomical Survey Unit. This publication makes use of data products from the Two Micron All Sky Survey, which is a joint project of the University of Massachusetts and the Infrared Processing and Analysis Center/California Institute of Technology, funded by the National Aeronautics and Space Administration and the National Science Foundation. This work has made use of data from the European Space Agency (ESA) mission
{\it Gaia} (\url{https://www.cosmos.esa.int/gaia}), processed by the {\it Gaia}
Data Processing and Analysis Consortium (DPAC,
\url{https://www.cosmos.esa.int/web/gaia/dpac/consortium}). Funding for the DPAC
has been provided by national institutions, in particular the institutions
participating in the {\it Gaia} Multilateral Agreement.
This research has made use of the VizieR catalogue access tool, CDS, Strasbourg, France. The original description of the VizieR service was published in A\&AS 143, 23.

\bibliography{biblio.bib} 
\bibliographystyle{mnras}


%
%
%
\bsp	
\label{lastpage}
\end{document}